\documentclass[11pt]{article}
\usepackage{xspace}
\usepackage{graphicx}
\usepackage{amsmath}
\usepackage{amssymb}
\usepackage{color}

\definecolor{Red}{rgb}{1,0,0}
\definecolor{Green}{rgb}{0,1,0}
\definecolor{Blue}{rgb}{0,0,1}
\definecolor{Black}{rgb}{0,0,0}

\def\beq{\begin{equation}}
\def\eeq#1{\label{#1}\end{equation}}
\def\eeqn{\end{equation}}

\def\beqa{\begin{eqnarray}}
\def\eeqa#1{\label{#1}\end{eqnarray}}
\def\eeqan{\end{eqnarray}}

\let\bar=\overbar

\def\Dslash{\not{\hbox{\kern-4pt $D$}}}
\def\dslash{\not{\hbox{\kern-2pt $\del$}}}

\def\msb{{\bar{\ssstyle M \kern -1pt S}}}

\def\sin22th{\textrm{sin}^2 2 \theta}
\def\dm2{\Delta m^2}
\def\Title#1{\begin{center} {\Large {\bf #1} } \end{center}}

\begin{document}

\Title{Searches for sterile
neutrinos using the T2K off-axis near detector}

\bigskip\bigskip


\begin{raggedright}  

{\it Debra Dewhurst on behalf of the T2K Collaboration\index{Dewhurst, D.},\\
Department of Physics\\
University of Oxford\\
OX1 3RH Oxford, UK}\\

\end{raggedright}
\vspace{1.cm}

{\small
\begin{flushleft}
\emph{To appear in the proceedings of the Prospects in Neutrino Physics Conference, 15 -- 17 December, 2014, held at Queen Mary University of London, UK.}
\end{flushleft}
}

\section{Introduction}

The results from a number of short baseline (SBL) neutrino experiments~\cite{Kaether:2010ag,Cheng:2012yy} 
and the reanalysis of previous reactor experiments with updated antineutrino fluxes~\cite{Mueller:2011nm}
suggest some incompatibility with the standard three-neutrino model (the gallium and reactor anomalies). A possible solution 
is the existence of sterile neutrinos~\cite{Acero:2007su}: 
right-handed particles that do not interact via the weak interaction. Their
existence can be studied through their mixing with the three active Standard Model neutrinos. The 3+1 model 
assumes there is only one sterile neutrino whose mixing is described by a unitary $4 \times 4$ 
matrix. If the mass squared difference $\Delta m^{2}_{41}$
is much larger than the other mass differences ($\mathcal{O}~1~\textrm{eV}^{2}$) the mixing can lead to 
SBL oscillations. The survival probability is given by
\def\beq{\begin{equation}}
$$P(\nu_{\alpha} \rightarrow \nu_{\alpha}) = 1 - {\sin22th}_{\alpha \alpha} \left(\frac{1.27 {\dm2}_{41} L_{\nu}}{E} \frac{[\textrm{GeV}]}{[\textrm{eV}]^{2}[\textrm{km}]}\right) $$
\def\eeqn{\end{equation}}
where $L_{\nu}$ and E are the flight path and energy of the neutrino respectively.

The existence of sterile neutrinos can be probed with the T2K Experiment~\cite{Abe:2011ks}: a long baseline 
neutrino oscillation experiment in Japan. 
Protons impinge on a
graphite target, producing a beam of $90 \%$ $\nu_{\mu}$, $8.8 \%$ $\bar{\nu_{\mu}}$, $1.1 \%$ $\nu_{e}$ and 
$0.1 \%~\bar{\nu_{e}}$. Neutrinos are detected at a near detector complex 280 m from 
the target, and at the Super-Kamiokande far detector 295 km from the target. The near detector complex consists 
of two detectors, one situated on-axis (INGRID) and one situated 
$2.5^{\circ}$ off-axis (ND280). At ND280 the $\nu_{\mu}$ component 
of the beam is peaked at 600~MeV/c and the dominant interaction is charged current (CC) quasi-elastic (QE) 
scattering ($\nu_{l} n \rightarrow l^{-} p$).
At higher energies pions are produced in CC resonant single pion production (CCRES), coherent pion production (CCCoh)
and multi pion production due to deep inelastic scattering (CCDIS).

Here we present a search for $\nu_{e}$ disappearance 
using ND280. The data analysed 
corresponds to an exposure of $5.9 \times 10^{20}$ protons on target (POT). We also present an 
introduction to a new analysis looking for $\nu_{\mu}$ disappearance with ND280, which promises to have 
interesting results in 2015.

\section{Search for short baseline oscillations using ND280}
\label{method}

To search for SBL oscillations, binned templates are built using the Monte Carlo (MC) reconstructed
energy distribution assuming CCQE interactions. The templates can be weighted event-by-event with the
oscillation probability to determine the dependence on the oscillation parameters of a given model. The oscillation
probability affects the signal events based on the true energy and flight path of the neutrino. 
These templates are then compared to data in a binned likelihood ratio fit 
with systematic errors included as nuisance parameters with gaussian constraints, similar to other T2K 
analyses~\cite{Abe:2013hdq}.  

\subsection{Search for short baseline $\nu_{e}$ disappearance}
\label{nuedisapp}

At ND280 the 3+1 model is investigated with $U_{\mu 4}=0$ in order to investigate the gallium and reactor anomalies.
A sample of $\nu_{e}$ events is selected with a purity (efficiency) of $63 \%$ (26$\%$)~\cite{Abe:2014usb}.
The largest background comes
from CCDIS or neutral current interactions where a $\pi^{0}$ is produced ($\nu_{\mu} N \rightarrow \pi^{0} X$). 
A control sample is used to measure this background, predominantly consisting of photon conversions from
$\nu_{\mu} N \rightarrow \pi^{0} X$ in neutral current and CCDIS interactions, with a purity (efficiency) of $92 \%$ ($12 \%$).

A measurement of $\nu_{\mu}$ CC 
interactions at ND280 is used to reduce the flux and the correlated cross-section uncertainties, 
as described in~\cite{Abe:2013hdq}. 
The sample of $\nu_{\mu}$ CC interactions is subdivided into events without charged pions (CC0$\pi$),
events with one positive pion (CC$\pi^{+}$) and other interactions that produce pions (CCOth).
This provides sensitivity to the rate of $\nu_{\mu}$ CCQE, CCRES and CCDIS interactions.

\begin{figure}[!ht]
\begin{center}
\includegraphics[width=0.45\columnwidth]{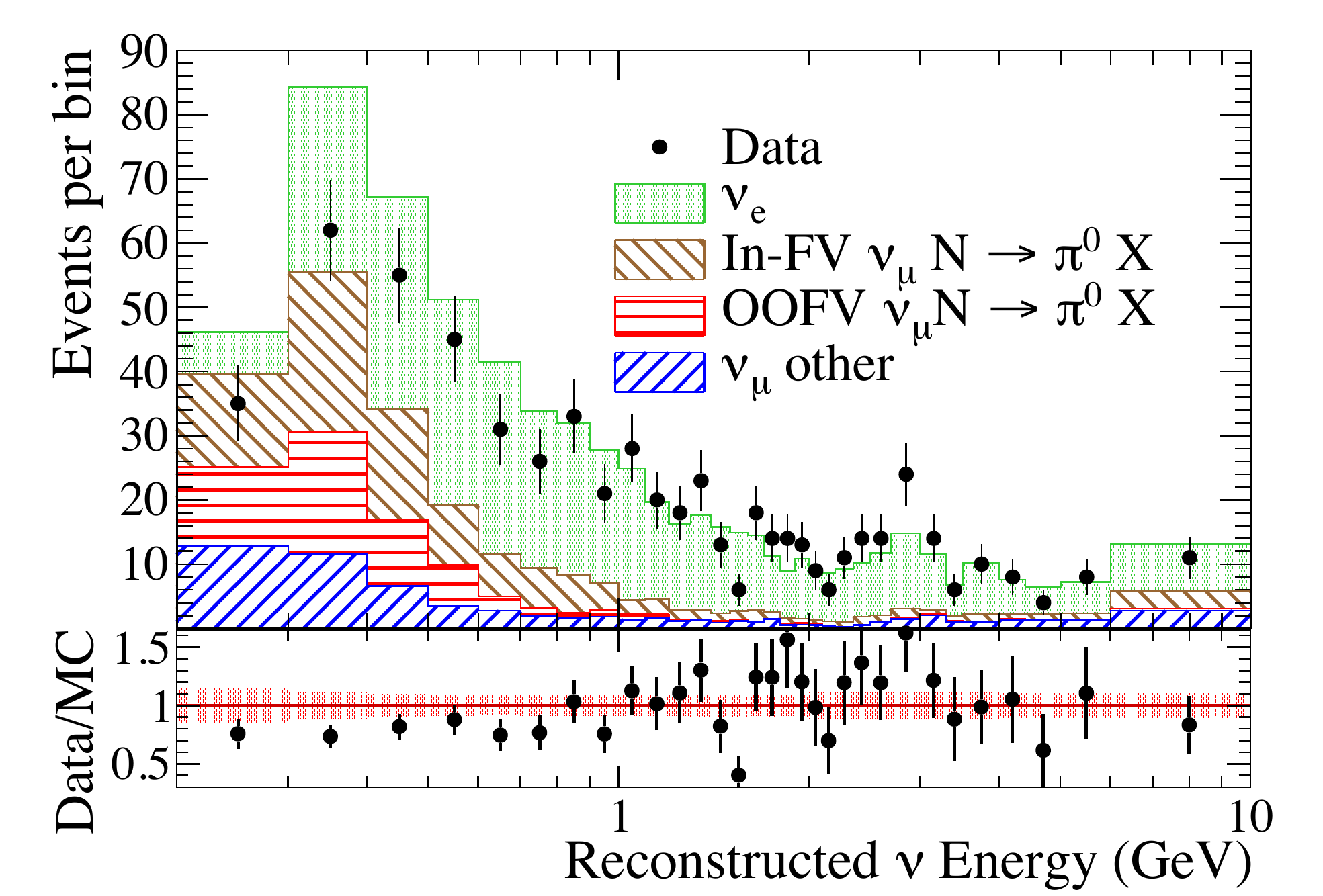}
\includegraphics[width=0.45\columnwidth]{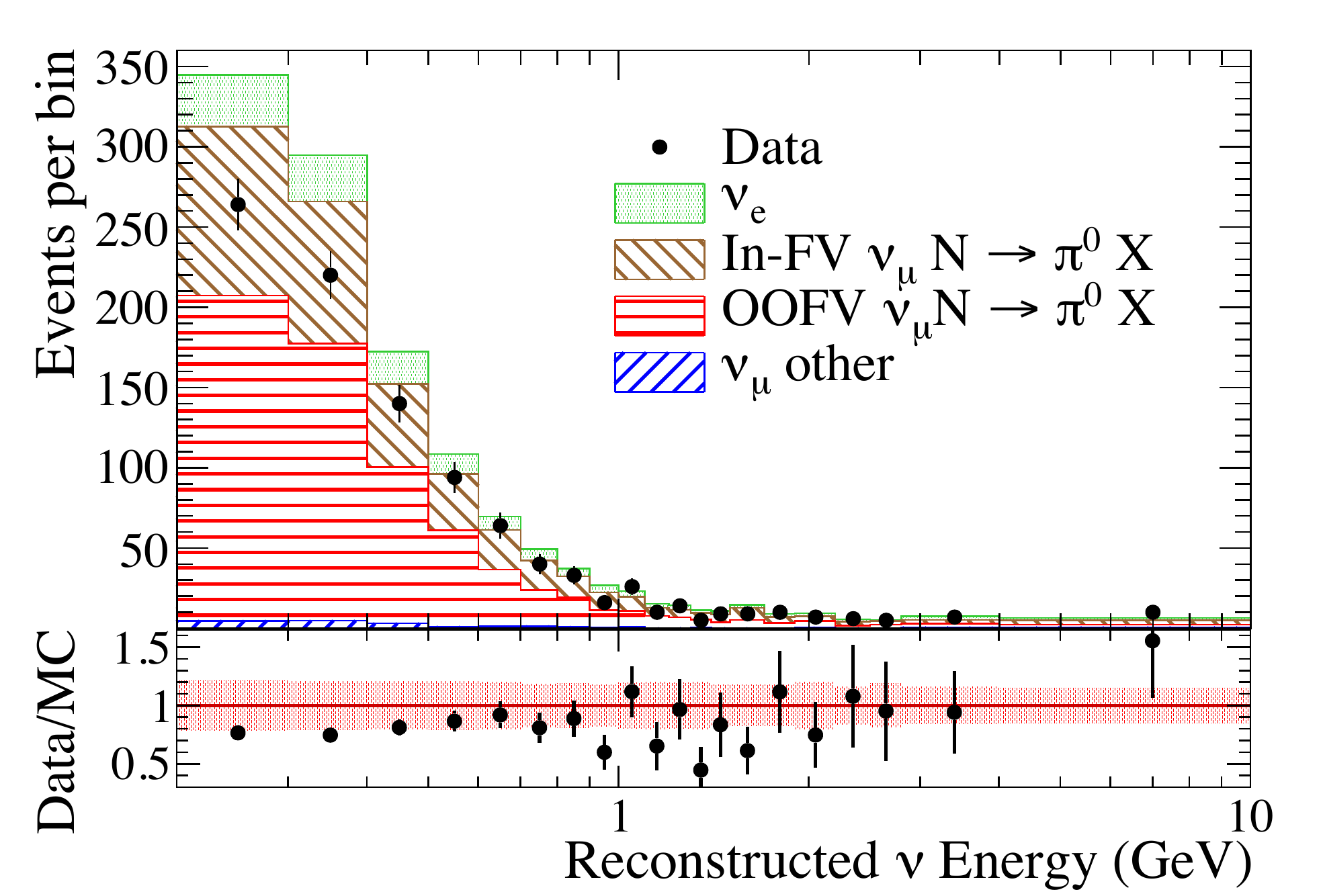}
\caption{Reconstructed energy distributions of the $\nu_{e}$ (left) and control (right) samples broken down
by $\nu_{e}$ interactions (signal), backgrounds inside and outside the fiducial volume due to $\nu_{\mu} N \rightarrow \pi^{0} X$ (In-FV and OOFV respectively),
and 
all other sources of background ($\nu_{\mu}$ other). The ratio of data
to MC in the null oscillation hypothesis is shown. The red error band corresponds to the fractional
systematic uncertainty. Black dots represent the data with statistical uncertainty.}
\label{fig:nue_reconstructed_energy}
\end{center}
\end{figure}

\begin{figure}[!ht]
\begin{center}
\includegraphics[width=0.5\columnwidth]{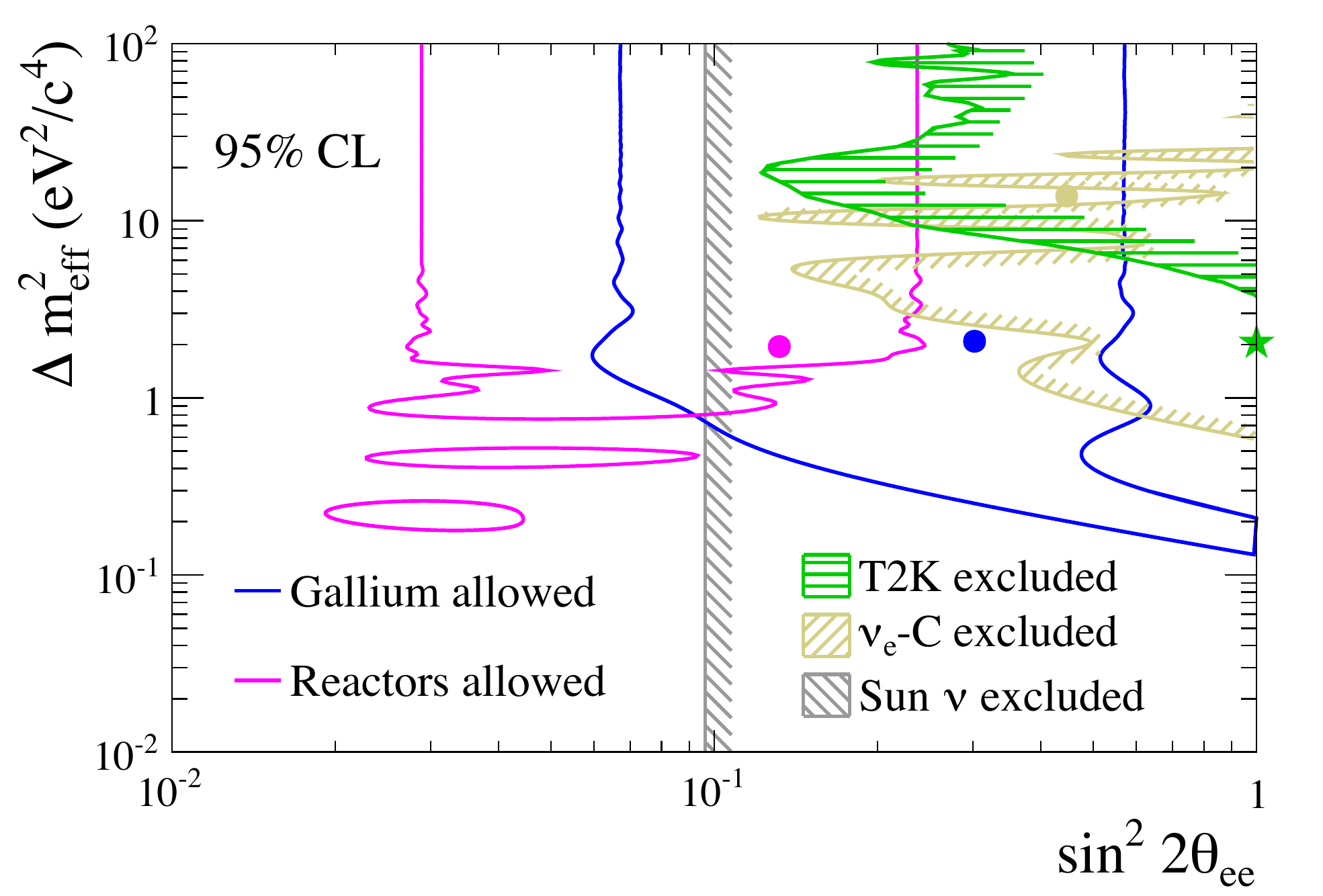}
\caption{The T2K exclusion region for $\nu_{e}$ disappearance at $95\%$ CL compared with 
other experimental results: allowed regions of gallium and reactor anomalies and excluded regions by $\nu_{e}$-
carbon interaction data and solar neutrino data~\cite{Giunti:2012tn}. The T2K best fit is marked by a green star, and those of other
experiments by filled circles of the same colour as the corresponding limits.}
\label{fig:nue_exclusion}
\end{center}
\end{figure}

The results from this analysis are based on data taken from January 2010-May 2013 (corresponding to $5.9 \times 10^{20}$ POT). 
Figure~\ref{fig:nue_reconstructed_energy} shows the reconstructed energy 
distributions of the $\nu_{e}$ 
signal and control samples. From the likelihood fit to data the best fit oscillation parameters are ${\sin22th}_{ee} = 1$ and 
${\dm2}_{\textrm{eff}} = 2.05~\textrm{eV}^{2}/\textrm{c}^{4}$~\cite{Abe:2014nuo}. The 2D confidence intervals in the 
${\sin22th}_{ee}$-${\dm2}_{\textrm{eff}}$ parameter space can be seen in Figure~\ref{fig:nue_exclusion}. 
The p-value of the null hypothesis is 0.085.

\subsection{Search for short baseline $\nu_{\mu}$ disappearance}

The signal sample for this analysis is the same as the $\nu_{\mu}$ CC sample described in Section~\ref{nuedisapp}, binned in terms
of the reconstructed neutrino energy assuming CCQE interactions and input into 
the likelihood fit described above. 

\begin{figure}[!ht]
\begin{center}
\includegraphics[width=0.5\columnwidth]{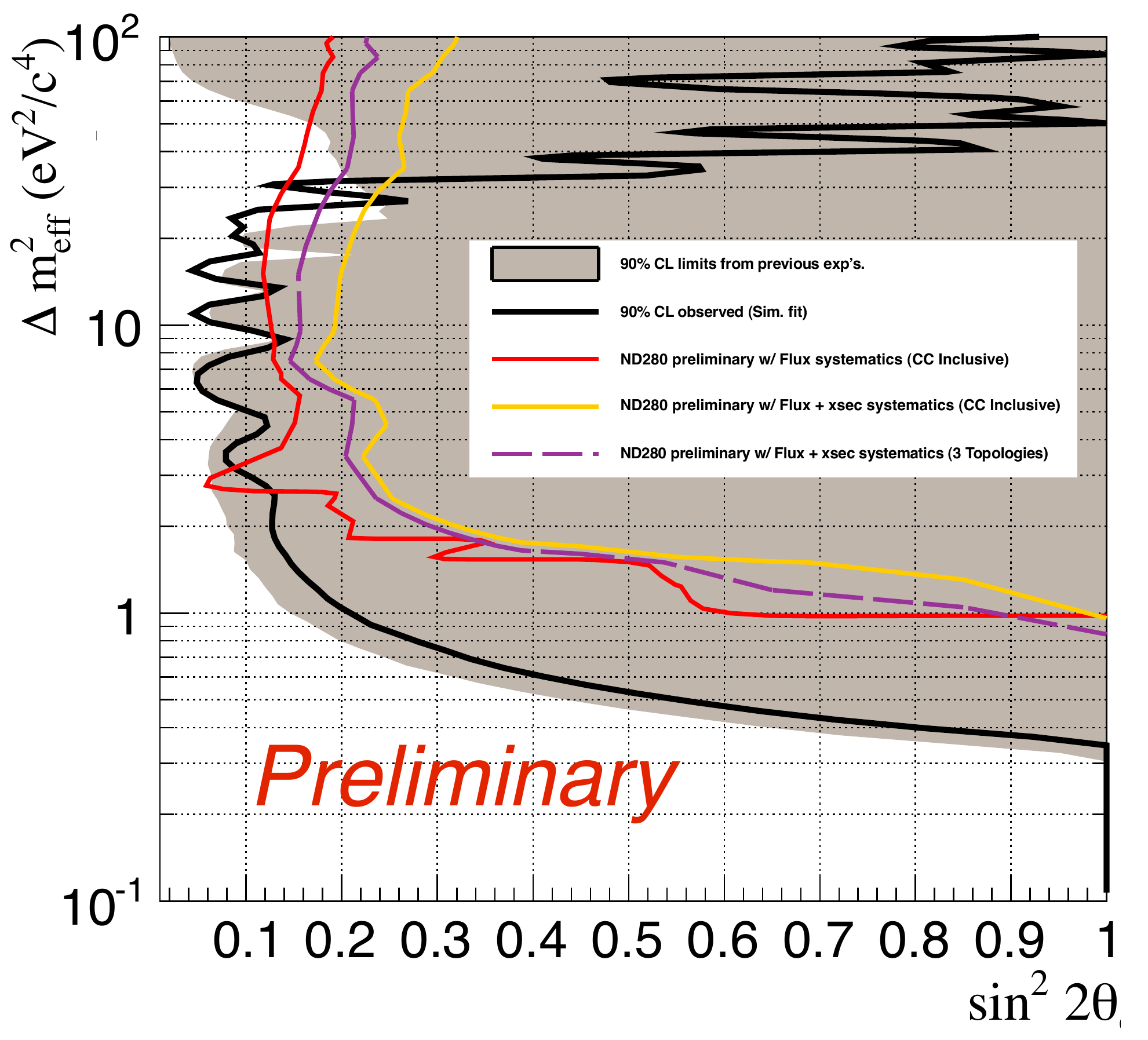}
\caption{The expected sensitivity for $\nu_{\mu}$ disappearance, at $90\%$ CL, based on $3\times 10^{20}$ POT of 
MC scaled to $6\times 10^{20}$ POT with flux and cross-section systematics included. The red and the
yellow lines show the $90\%$ CL when the CC0$\pi$, CC$\pi^{+}$ and CCOth samples are combined into a single CC inclusive sample. The dashed purple line
shows the $90\%$ CL when the three samples are kept
separate. The shaded region
indicates the 90$\%$ CL limits from the CCFR~\cite{Stockdale:1984cg} and CDHS~\cite{Dydak:1983zq} experiments. The black line represents 
the $90\%$ CL limits from MiniBooNE/SciBooNE measurements~\cite{Mahn:2011ea}.}
\label{fig:numu_sensitivity}
\end{center}
\end{figure}

Figure~\ref{fig:numu_sensitivity} shows the 
expected sensitivity for a 3+1 analysis, at $90\%$ CL, when flux and cross-section systematics are evaluated, compared 
to other experimental results. 
These preliminary results are promising. Once the detector systematics
and final state interaction systematics have been included the data will be analysed.

\section{Summary}

A search for $\nu_{e}$ disappearance caused by SBL oscillations has been performed with the T2K off-axis
near detector. The 
exclusion region at $95 \%$ CL is approximately given by ${\sin22th}_{ee} > 0.3$ and 
${\dm2}_{\textrm{eff}} >7~\textrm{eV}^{2}/\textrm{c}^{4}$. The p-value of the null oscillation hypothesis is 0.085. 
These results exclude parts
of the gallium anomaly and a small part of the rector anomaly allowed regions. The analysis is limited
by statistical uncertainties and therefore further data from T2K will help to improve the analysis.

Searches for $\nu_{\mu}$ disappearance due to SBL oscillations are being constructed. A full
MC sensitivity study with systematic uncertainties is almost complete for the 3+1 model and promises
to have interesting results in 2015.

\end{document}